\documentclass[10pt,letterpaper,compsoc,conference]{iiswc26}

\usepackage{cite}
\usepackage{amsmath,amssymb,amsfonts}
\usepackage{algorithmic}
\usepackage{graphicx}
\usepackage{array}
\usepackage[dvipsnames]{xcolor}
\usepackage[final]{microtype}
\usepackage[italic]{mathastext}
\usepackage{libertine}
\usepackage[T1]{fontenc}
\usepackage{textcomp}
\usepackage[varqu,varl]{zi4}
\usepackage[all]{nowidow}
\usepackage{flushend}

\usepackage{comment}

\graphicspath{{figs/}}

\IEEEoverridecommandlockouts

\begin{document}

\title{Activation Concentration: Characterizing Column-Level Output Sparsity Across Diffusion Model Architectures}

\author{
\IEEEauthorblockN{Dazhi Yang\IEEEauthorrefmark{1},
Shafayat Mowla Anik\IEEEauthorrefmark{2},
Byeong Kil Lee\IEEEauthorrefmark{2},
Jeeho Ryoo\IEEEauthorrefmark{1}}
\IEEEauthorblockA{\IEEEauthorrefmark{1}Fairleigh Dickinson University, Vancouver, BC, Canada\\
Email: d.yang@student.fdu.edu, j.ryoo@fdu.edu}
\IEEEauthorblockA{\IEEEauthorrefmark{2}University of Colorado at Colorado Springs, Colorado Springs, CO\\
Email: \{sanik, blee\}@uccs.edu}
\thanks{This work has been submitted to the IEEE for possible publication. Copyright may be transferred without notice, after which this version may no longer be accessible.}
}

\maketitle
\pagestyle{plain}

\begin{abstract}
Recent diffusion accelerators exploit activation sparsity by skipping near-zero GELU outputs, reporting 52--85\% element-level sparsity.  However, systolic-array hardware processes activations at column granularity, where a single non-zero element forces the entire column to be computed.  We present the first systematic column-level sparsity characterization across seven diffusion workloads spanning three workload groups and four modalities.  Our measurements reveal that element-level sparsity overstates hardware-exploitable sparsity by up to 78 percentage points and exposes a three-way taxonomy.  UNet+transformer workloads exhibit activation concentration with workload-dependent cycle reductions up to 30.6\%.  Pure-transformer DiT shows dispersion, yielding 12.4\%.  Motion/dance transformer workloads range from modest reductions to 50.8\% for MLD, driven by its extreme token dimension and expansion ratio.  Cycle-level simulation on a GDDR6-based accelerator confirms that memory stalls account for up to 84--89\% of total cycles and that layout sensitivity tracks the profiling-based taxonomy.  A full accuracy sweep across five thresholds reveals that UNet+transformer workloads degrade gracefully, while motion models exhibit an accuracy cliff between the primary operating point and the next threshold.  Our characterization shows that workload group and model dimensions jointly determine whether column-level memory layout optimization is beneficial, and element-level sparsity alone is insufficient for that prediction.
\end{abstract}

\section{Introduction}
\label{sec:introduction}

Diffusion models have become the dominant generative framework across image~\cite{rombach2022latent}, video~\cite{chen2024videocrafter2}, audio~\cite{huang2023makeanaudio}, and motion~\cite{tevet2023mdm} synthesis.  These models differ substantially in their neural-network backbones: image generators embed transformer blocks inside UNet encoder-decoders with spatially varying token dimensions, pure transformer architectures such as DiT~\cite{peebles2023dit} process a fixed token sequence through uniform layers, and motion or dance models operate on compact skeletal representations with as few as six tokens per sequence.  Despite this architectural diversity, each model's feed-forward network (FFN) layers share a common $\text{fc}_1 \rightarrow \text{GELU} \rightarrow \text{fc}_2$ structure whose near-zero GELU outputs create activation sparsity that hardware can exploit.  Prior work reports 52--85\% element-level sparsity across three diffusion workloads~\cite{heo2025exion}, suggesting that this optimization opportunity is broadly and uniformly available.

However, this element-level metric does not reflect hardware reality.  Systolic-array accelerators do not process individual elements in isolation but instead tile activations at column granularity, and a single non-zero element anywhere in a column forces the entire column to be computed.  Column-level sparsity, defined as the fraction of columns that are entirely zero, is therefore the hardware-relevant metric, and it is always less than or equal to element-level sparsity.  The gap between the two quantifies how much element-level measurements overstate the sparsity actually exploitable by tiled hardware.  Moreover, existing characterization covers only three workloads from two modalities (image and video), all sharing high token dimensions and similar FFN configurations, leaving open the question of whether the reported sparsity levels generalize to the broader and more structurally diverse landscape of production diffusion models.

\begin{figure}[t]
  \centering
  \includegraphics[width=\columnwidth]{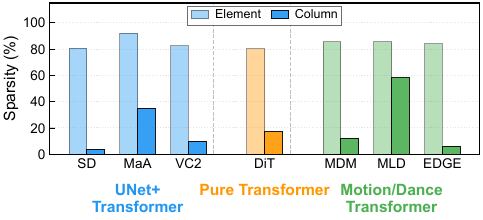}
  \caption{Element vs.\ column sparsity per model}
  \label{fig:teaser}
\end{figure}

We present the first systematic column-level sparsity characterization across seven diffusion workloads spanning three workload groups: UNet+transformer workloads (Stable Diffusion, VideoCrafter2, Make-an-Audio), the pure transformer workload DiT, and motion/dance transformer workloads (MDM, MLD, EDGE).  Figure~\ref{fig:teaser} compares element-level sparsity (light bars) and column-level sparsity (dark bars) at threshold $\tau{=}0.164$ for all seven models.  The light bars are uniformly high (52--85\%), but the dark bars reveal that column sparsity varies from as low as 6.0\% to 58.3\%, exposing a three-way workload-regime split that element-level analysis entirely obscures.  The gap between the two metrics reaches up to 78 percentage points for motion/dance transformer workloads, where nearly every column contains at least one non-zero element despite high scalar sparsity.  Across denoising iterations, the three workload groups also exhibit distinct temporal behaviors that determine whether a static memory layout is viable.  UNet+transformer workloads exhibit activation concentration: the set of hot columns stabilizes after the first denoising iteration and remains largely unchanged, meaning a one-time layout decision is structurally sound.  The pure transformer DiT instead exhibits dispersion, where column sparsity declines from ${\sim}$43\% to ${\sim}$10\% over 50 iterations as the model recruits additional features during refinement, though the hot-column identity never changes because columns only turn on, never off.  Motion/dance transformer workloads present a mixed picture: MDM and EDGE show moderate column sparsity with high temporal stability, while MLD combines the highest column sparsity of any model (58.3\%) with the lowest stability, driven by its extreme token dimension ($M{=}6$) and $4\times$ expansion ratio.

These characterization findings translate to measurable differences in memory-system performance.  The characterization on cycle-accurate simulator shows that memory stalls account for 84--89\% of total execution cycles across all seven models under a baseline row-major layout.  A hot-cold memory layout that groups cold columns contiguously to improve DRAM row-buffer locality yields cycle reductions ranging from 3.9\% (Stable Diffusion) to 50.8\% (MLD), tracking the profiling-based taxonomy: models with high column sparsity and stable temporal patterns benefit the most, while models with low column sparsity see minimal improvement regardless of their element-level sparsity.  A full accuracy sweep across five thresholds further reveals that motion models exhibit a sharp accuracy cliff between $\tau{=}0.164$ and $\tau{=}0.17$, where quality degrades abruptly, while UNet+transformer and pure transformer workloads degrade more gracefully across the same range.  This paper makes three contributions.
\begin{enumerate}
\item The first column-level sparsity characterization across seven diffusion workloads and three workload groups, revealing that element sparsity overstates exploitable sparsity by up to 78 percentage points.

\item Identification of activation concentration and dispersion as workload-dependent temporal phenomena that govern static memory layout viability.

\item A three-way taxonomy mapping workload group and model dimensions to layout optimization amenability, with cycle reductions ranging from 3.9\% to 50.8\%.


\end{enumerate}

The rest of this paper is organized as follows.  Section~\ref{sec:background} provides background on diffusion architectures, FFN-Reuse, and DRAM row-buffer locality.  Section~\ref{sec:methodology} describes the profiling pipeline, workloads, threshold sweep, and cycle-level simulator.  Section~\ref{sec:characterization} presents the column-level sparsity characterization and three-way taxonomy.  Section~\ref{sec:evaluation} validates the taxonomy through cycle-level simulation and accuracy evaluation.  Section~\ref{sec:discussion} discusses implications and limitations, Section~\ref{sec:related} surveys related work, and Section~\ref{sec:conclusion} concludes.

\section{Background and Motivation}
\label{sec:background}

\subsection{Diffusion Model Inference}

Diffusion models generate outputs through an iterative denoising process.  Starting from Gaussian noise, the model applies a learned denoiser for $T$ steps (typically 50) to progressively refine the output.  Each denoising step involves a forward pass through the model's neural network, which contains feed-forward network (FFN) blocks with the structure $\text{fc}_1 \rightarrow \text{GELU} \rightarrow \text{fc}_2$.  The GELU activation function produces outputs that are often near-zero, creating activation sparsity that hardware can potentially exploit.

\begin{figure}[t]
  \centering
  \includegraphics[width=\columnwidth]{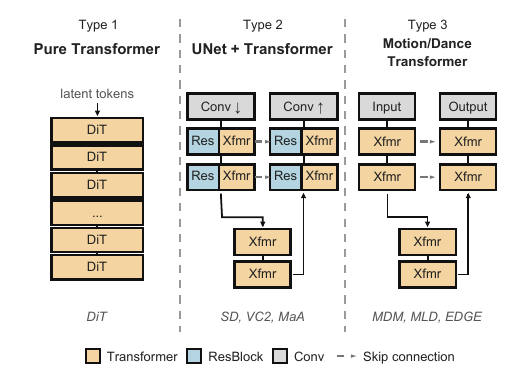}
  \caption{Three workload groups used in this paper.}
  \label{fig:arch_types}
\end{figure}

We organize the seven workloads into three groups that capture the FFN structures and sparsity regimes evaluated in this paper.  Figure~\ref{fig:arch_types} depicts the groups side by side.  The left column shows the pure transformer workload DiT~\cite{peebles2023dit}, which stacks transformer blocks with multi-head self-attention and FFN layers, processing all tokens uniformly with a token dimension $M$ that stays constant across layers.  The center column shows UNet+transformer workloads (Stable Diffusion~\cite{rombach2022latent}, VideoCrafter2~\cite{chen2024videocrafter2}, Make-an-Audio~\cite{huang2023makeanaudio}), which embed transformer blocks inside a UNet encoder-decoder so that $M$ varies across resolution levels.  The right column shows motion/dance transformer workloads (MDM~\cite{tevet2023mdm}, MLD~\cite{chen2023mld}, EDGE~\cite{tseng2023edge}), which use transformer-based motion or dance denoisers with distinct token dimensions and expansion ratios.  These groups differ in token dimension ($M$), hidden dimension ($N$), expansion ratio, and temporal stability, parameters that, as we show, determine column-level sparsity behavior.

\begin{figure}[h]
  \centering
  \includegraphics[width=\columnwidth]{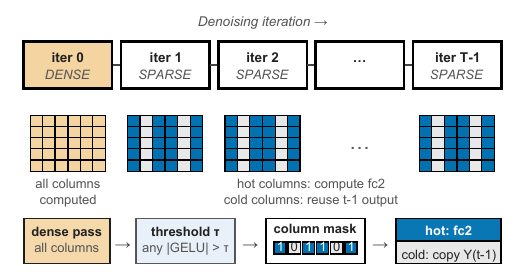}
  \caption{FFN-Reuse dataflow.}
  \label{fig:ffn_reuse}
\end{figure}

\subsection{Output Sparsity and FFN-Reuse}

The FFN-Reuse optimization~\cite{heo2025exion} exploits a key observation.  After the first denoising iteration, many GELU outputs remain near-zero across subsequent iterations.  Figure~\ref{fig:ffn_reuse} organizes the mechanism from top to bottom.  The top row separates the dense bootstrap iteration from later sparse denoising iterations.  The middle row shows the corresponding column activity.  Iteration~0 computes all $N$ columns, whereas later iterations recompute only hot columns and leave cold columns gray.  The bottom pipeline shows how the sparse decision is made.  A dense pass exposes GELU activations, a threshold $\tau$ converts them into a column mask, and the accelerator computes $\text{fc}_2$ only for columns whose mask bit is one.  Hot columns perform the $\text{fc}_2$ multiply and fetch $\text{W}_2$, while cold columns skip both operations and reuse $Y(t{-}1)$.

This optimization is column-granular on a systolic-array accelerator, so it skips entire column tiles rather than individual scalar activations.  A column $j$ is therefore classified as hot if any token exceeds the threshold ($\exists\, i \text{ such that } |\text{GELU output}_{ij}| > \tau$), and cold otherwise.  The potential speedup depends on two factors, how many columns are cold and how the remaining hot columns are arranged in memory.  Prior work reports 52--85\% element-level sparsity across DiT, Stable Diffusion, and VideoCrafter2~\cite{heo2025exion}, suggesting that the optimization opportunity is large at the element level.  Our profiler reproduces these element-level numbers as a sanity check before evaluating whether they translate to exploitable column-level sparsity.

\subsection{Column-Level Granularity}
\label{sec:bg_column}

The distinction between element-level and column-level sparsity is critical for hardware designers.  In a systolic-array accelerator, the processing unit tiles the activation matrix $\mathbf{A} \in \mathbb{R}^{M \times N}$ along the $N$ (column) dimension.  Each column corresponds to one hidden-dimension output, and the $M$ rows correspond to tokens (spatial positions, time steps, or batch elements).  If any of the $M$ tokens in a column has a non-zero activation, the accelerator must compute the entire column because it cannot skip individual elements within a column tile.

\begin{figure}[t]
  \centering
  \includegraphics[width=\columnwidth]{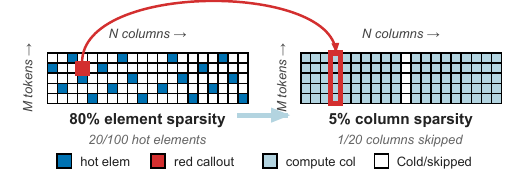}
  \caption{Element- vs.\ column-level sparsity.}
  \label{fig:concept}
\end{figure}

Column-level sparsity is therefore always $\leq$ element-level sparsity.  Figure~\ref{fig:concept} shows the same $M \times N$ activation matrix viewed at two granularities.  The left panel displays the element-level view, where each cell represents one scalar activation and cold cells (below threshold) are shaded gray.  Most individual elements are cold, producing high element-level sparsity.  The right panel displays the column-level view, where the hardware decision is made per column rather than per element.  Each column is shaded based on whether it contains any hot element.  The highlighted red element in one column is enough to make that entire column hot, so the accelerator must still compute it despite the remaining elements being cold.  This example shows why 80\% element sparsity can collapse to only 5\% column sparsity.  The size of this gap depends on the token dimension $M$, because larger $M$ gives each column more chances to contain at least one hot element, making column sparsity increasingly harder to achieve.

To build intuition, consider $M$ independent tokens per column, each with probability $p$ of falling below threshold.  The probability that the entire column is cold is $p^M$, so column-level sparsity scales as $p^M$ while element-level sparsity remains $p$ regardless of $M$.  Even moderately high per-element sparsity ($p{=}0.85$) yields $0.85^{256} \approx 10^{-18}$ at $M{=}256$, effectively zero column sparsity, whereas $0.85^{6} \approx 0.38$ at $M{=}6$ preserves a substantial fraction.  The seven workloads in this study span $M{=}6$ (MLD) to over 10{,}000 (VideoCrafter2), so the independence assumption is only a first-order approximation because tokens within the same layer share attention context and are not statistically independent.  Nonetheless, this exponential scaling explains why models with small $M$ and high expansion ratios can sustain column sparsity that models with large $M$ cannot.

\subsection{DRAM Row-Buffer Locality}

The accelerator's off-chip memory system uses GDDR6 DRAM organized into channels and banks, each with a row buffer.  Accessing data within an already-open row (a row hit) is significantly faster than opening a new row (a row miss or row conflict).  Row conflicts, in which a new row must be opened in the same bank after precharging the current row, are particularly costly, adding tens of nanoseconds per access.  The physical layout of activation data in DRAM therefore determines which rows are accessed during computation and how frequently conflicts occur.

\begin{figure}[t]
  \centering
  \includegraphics[width=\columnwidth]{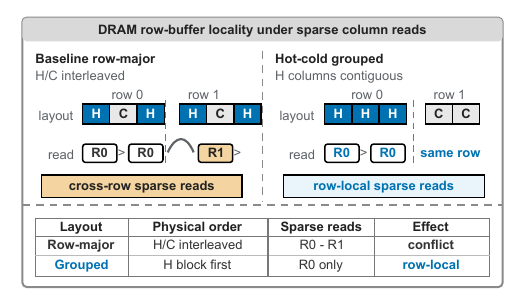}
  \caption{DRAM row-major and grouped layouts.}
  \label{fig:dram_layout}
\end{figure}

Figure~\ref{fig:dram_layout} shows how layout affects row-buffer behavior when some columns are skipped.  The left half shows a row-major layout where columns are stored in their original order.  When a sparse execution path reads only a subset of columns, the accessed columns may fall in different DRAM rows, causing row switches.  The right half shows an alternative layout where the columns that are read together are grouped contiguously in memory.  In this example, the same set of reads now falls within fewer rows, reducing row-buffer conflicts.  Whether such grouping is feasible for a given workload depends on whether the activation pattern produces enough cold columns at hardware granularity and whether those columns remain consistent across denoising iterations.  These are open questions that motivate the characterization study in Section~\ref{sec:characterization}.

\section{Methodology}
\label{sec:methodology}

\subsection{Profiling Pipeline}

We instrument each model's GELU activations using PyTorch forward hooks registered on \texttt{nn.GELU} modules, supplemented by a monkey-patch on \texttt{F.gelu} for models that call the functional API directly.  For models using GEGLU variants (Stable Diffusion, VideoCrafter2, Make-an-Audio), we additionally hook the gating module to capture the full activation tensor.  For each FFN layer at each denoising iteration, the profiler records a full-precision activation bitmask.  A column is classified as hot if any of its $M$ token activations exceeds the threshold $\tau$ in absolute value, and cold otherwise; no sampling or approximation is applied, and every element is evaluated.  To quantify temporal stability, we compute the Jaccard similarity between the hot-column sets of consecutive iterations, defined as the size of their intersection divided by the size of their union.  A Jaccard index of 1.0 indicates that the hot-column set is identical across iterations, justifying a static memory layout, while lower values indicate that columns are switching between hot and cold states.


\subsection{Workloads}

Table~\ref{tab:workloads} summarizes the seven diffusion models we profile, spanning three workload groups and four modalities (image, video, audio, motion).  We select these workloads to cover the design space of FFN configurations encountered in production diffusion pipelines, from $M{=}6$ tokens per column in MLD to over 10{,}000 in VideoCrafter2, and from $N{=}1{,}024$ to $N{=}5{,}120$ hidden dimensions.  The spread is deliberate because small $M$ stresses the column-statistical argument developed in Section~\ref{sec:m_dimension}, while large $M$ and $N$ exercise the systolic-array and DRAM behavior modeled by the cycle-level simulator.  The expansion-ratio range (2$\times$ and 4$\times$, including GEGLU gating that doubles $\text{fc}_1$) lets us isolate which model parameters drive column-level sparsity and which do not, extending rather than redefining the FFN-Reuse design space formulated by the state-of-the-art diffusion accelerator work~\cite{heo2025exion}.

\begin{table}[t]
\centering
\caption{Workload summary.}
\label{tab:workloads}
\small
\setlength{\tabcolsep}{3pt}
\renewcommand{\arraystretch}{0.95}
\begin{tabular}{@{}llrcccl@{}}
\hline
Model & Group & $L$ & $M$ & $N$ & Exp. & Mod. \\
\hline
DiT-XL/2 & Pure-Xfmr & 28 & 256 & 4608 & 4$\times$ & Img. \\
SD v1.4 & U+Xfmr & 16 & 256--4096 & 1280--5120 & 4$\times$ & Img. \\
VC2 & U+Xfmr & 33 & 2560--10240 & 1280--5120 & 4$\times$ & Vid. \\
MaA & U+Xfmr & 11 & 200--800 & 1280--2560 & 4$\times$ & Aud. \\
MDM & Mot-Xfmr & 8 & 242 & 1024 & 2$\times$ & Mot. \\
MLD & Mot-Xfmr & 9 & 6 & 1024 & 4$\times$ & Mot. \\
EDGE & Mot-Xfmr & 10 & 3300 & 1024 & 2$\times$ & Dance \\
\hline
\end{tabular}
\renewcommand{\arraystretch}{1.0}
\end{table}

\subsection{Threshold Sweep Design}

We evaluate the column-masking opportunity along two orthogonal axes.  The uniform axis sweeps a global activation magnitude threshold $\tau$, classifying a column as hot if any element exceeds $\tau$.  The per-layer axis sweeps a target hot ratio $r$ obtained by a layer-wise binary search that calibrates a separate threshold per layer.  The same five values (0.10, 0.15, 0.164, 0.17, 0.20) are used for both axes, but the semantics differ.  $\tau$ is a physical activation magnitude, while $r$ is the resulting fraction of hot columns after calibration.  The same numerical label (e.g., 0.164) therefore means $\tau{=}0.164$ in the uniform case but $r{=}0.164$ (target 16.4\% hot columns) in the per-layer case.  We anchor the analysis at $\tau{=}0.164$, the largest threshold from our preliminary sweep that sits below the sharp motion-quality cliff observed between $\tau{=}0.164$ and $\tau{=}0.17$.  For layers without durable natural sparsity (e.g., MDM and EDGE), the per-layer search pushes thresholds far above the uniform-sweep range, a phenomenon we term threshold inflation and analyse in Section~\ref{sec:perlayer_limits}. Threshold inflation occurs when per-layer calibration is forced to set thresholds well above the physical activation range in order to meet a target hot ratio, indicating that the layer lacks natural column sparsity at any meaningful threshold. 

\begin{table}[h]
\centering
\caption{Simulated system configuration.}
\label{tab:simconfig}
\small
\setlength{\tabcolsep}{2pt}
\renewcommand{\arraystretch}{1.02}
\begin{tabular}{@{}ll@{}}
\hline
\multicolumn{2}{c}{\textbf{Accelerator configuration}} \\
\hline
Core & compute/memory abstraction, 800\,MHz \\
Compute array & $16{\times}16$ systolic matrix engine \\
Element unit & 64-wide SIMD, 16-bit elements \\
SRAM buffers & input/weight/output = 24/192/24\,KB \\
Buffer slots & input/weight/output = 2/3/2 \\
\hline
\multicolumn{2}{c}{\textbf{DRAM configuration}} \\
\hline
DRAM & GDDR6 16-Gb x16 2000\,MT/s\\
Timing & CL/RD/WD/RP/RAS = 24/26/16/26/53 \\
Column timing & CCDS/CCDL = 4/6 \\
Channels & 6 channels, 64-bit/channel \\
Bandwidth & 96\,GB/s \\
Controller & FR-FCFS, open-row, refresh \\
Mapping & RoBaRaCoCh, 8KB row \\
\hline
\end{tabular}
\renewcommand{\arraystretch}{1.0}
\end{table}

\subsection{Accuracy Evaluation}

We evaluate output quality by comparing dense baseline outputs against column-masked outputs for all seven models at the primary operating point $\tau{=}0.164$.  For each model, we generate $N{=}100$ outputs under both the dense (all-column) and masked (cold-columns-zeroed) configurations using identical random seeds, so that any quality difference is attributable solely to the column masking. This paired setup measures masking-induced output shift rather than standalone generation quality. 

\subsection{Cycle-Level Simulator}
Following prior state-of-the-art diffusion accelerator work~\cite{heo2025exion}, we develop a custom cycle-level simulator and integrate it with Ramulator~2.0~\cite{luo2024ramulator2}, a widely used DRAM simulator, to model DRAM latency at the channel, bank, and row-buffer level.
The accelerator frontend models a $16{\times}16$ systolic matrix engine, a 64-wide SIMD elementwise unit, and separate on-chip input, weight, and output buffers.  These on-chip buffers provide 24\,KB, 192\,KB, and 24\,KB of capacity for input, weight, and output data, respectively, with input and output double-buffered and weight triple-buffered to hide data-fetching latency.  The simulator drives the FFN-Reuse dataflow on this model, consuming the column bitmasks recorded by the profiling pipeline and emitting per-iteration cycle counts decomposed into compute, memory-stall, and other components.  Table~\ref{tab:simconfig} lists the full hardware configuration.  Each run executes 50 denoising iterations against a per-column hot/cold bitmask.



\section{Column-Level Sparsity Characterization}
\label{sec:characterization}
\begin{figure}[t]
  \centering
  \includegraphics[width=\columnwidth]{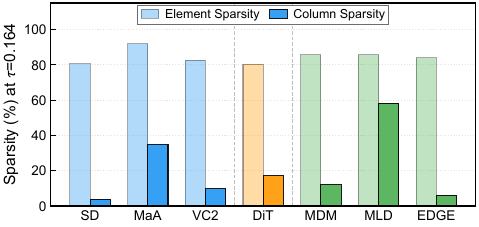}
  \caption{Element vs.\ column sparsity at $\tau{=}0.164$.}
  \label{fig:elem_vs_col}
\end{figure}

\subsection{Element vs.\ Column Sparsity and the Granularity Gap}
\label{sec:granularity_gap}

Figure~\ref{fig:elem_vs_col} compares element-level and column-level sparsity at $\tau{=}0.164$ for all seven models (iteration 1+ average), showing that the gap between the two metrics varies dramatically by workload group.  For UNet+transformer workloads, element sparsity of 81--92\% translates to column sparsity of 4--35\% (iter~1+ weighted average), with the exploitable fraction strongly workload-dependent despite high scalar sparsity.  For the pure transformer DiT, element sparsity of ${\sim}$80\% corresponds to ${\sim}$17\% average column sparsity, with the gap growing over iterations as more columns gain at least one non-zero element.  For motion/dance transformer workloads, MDM and EDGE exhibit 84--86\% element sparsity but only 12.1\% and 6.0\% column sparsity, respectively.  Element sparsity is therefore a misleading signal for these workloads because nearly every column contains at least one non-zero token.  MLD is the highest-sparsity case in this group at 58.3\%, which we explain in Section~\ref{sec:m_dimension}.

\subsection{Activation Concentration and Dispersion}
\label{sec:concentration}

\begin{figure}[t]
  \centering
  \includegraphics[width=\columnwidth]{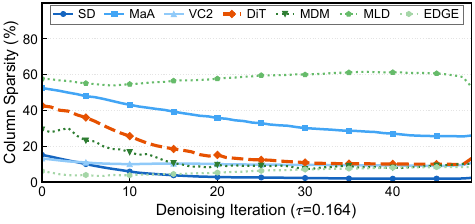}
  \caption{Column sparsity per iteration.}
  \label{fig:concentration}
\end{figure}

Figure~\ref{fig:concentration} plots column sparsity across denoising iterations for each model, which is the signature result of this study.  Three distinct temporal patterns emerge, each tied to workload structure and model dimensions.  \textbf{Concentration (UNet+transformer workloads).}  Stable Diffusion, Make-an-Audio, and VideoCrafter2 jump from the dense iteration to a sparse regime at iteration~1, reaching 14.1\%, 51.7\%, and 12.6\% column sparsity respectively, with mean Jaccard similarity (Section~\ref{sec:methodology}) of 0.91, 0.70, and 0.91 across consecutive sparse iterations, indicating that the identity of hot columns changes little from one iteration to the next.  The hot-column set is therefore largely set early, supporting a static hot-cold layout whose realized benefit still depends on workload-specific cold-column counts.  \textbf{Dispersion (pure transformer workload).}  DiT begins with ${\sim}$42.6\% column sparsity at iteration~0 and gradually declines to ${\sim}$10\% by iteration~40+.  Unlike the UNet+transformer pattern, more columns become active over time as the model recruits additional features during refinement.  Remarkably, however, the Jaccard index between consecutive iterations is also 1.0.  Columns only turn on, never off, so the cold set at any iteration is a strict subset of the cold set at iteration~0, and a static layout from iteration~0 remains valid throughout.  \textbf{Low-to-mixed (motion/dance transformer workloads).}  MDM and EDGE show column sparsity of 12.1\% and 6.0\% on average (iter~1+ weighted), still well below the strongest cases but not negligible.  MLD shows 58.3\% average column sparsity with the lowest temporal stability of any model (Jaccard mean 0.433), which we analyze next.

\subsection{The M-Dimension and Expansion Ratio Effect}
\label{sec:m_dimension}

\begin{figure}[t]
  \centering
  \includegraphics[width=\columnwidth]{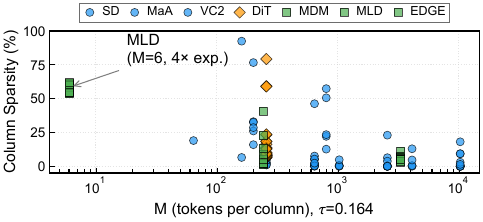}
  \caption{Column sparsity vs.\ token dimension $M$.}
  \label{fig:m_scatter}
\end{figure}

Figure~\ref{fig:m_scatter} relates per-layer column sparsity to the token dimension $M$ (the number of rows in the activation matrix), colored by workload group, to explain why some models retain cold columns while others do not.  Large $M$ for UNet+transformer workloads (200--10{,}240 tokens) decreases the probability that all $M$ tokens in a column are below threshold, because each column has more chances to contain at least one hot token.  However, $M$ alone is not sufficient.  EDGE has $M{=}3{,}300$ yet shows low column sparsity because its observed activation distribution is more dispersed across columns, while MLD has very small $M$ and a wider expansion ratio that create many columns that remain entirely cold.

The MLD exception is explained by two compounding factors.  First, MLD has the smallest token dimension of any model ($M{=}6$), meaning only six tokens must all be below threshold for a column to be cold, which is a much easier condition.  Second, MLD's expansion ratio is $4\times$ ($\text{hidden\_dim}{=}1024$, $\text{d\_model}{=}256$), compared to $2\times$ for MDM and EDGE ($\text{hidden\_dim}{=}1024$, $\text{d\_model}{=}512$).  The wider expansion creates more redundant columns in the hidden space, increasing the likelihood that some are entirely inactive.  This combination of extreme $M$ and high expansion ratio explains why MLD achieves 58.3\% column sparsity while MDM and EDGE remain much lower despite belonging to the same motion/dance workload group.

\subsection{Per-Layer Variation and Calibration Limits}
\label{sec:perlayer_limits}

Column sparsity varies substantially across layers within a single model.  Stable Diffusion ranges from 0.1\% to 19.1\% across its 16 transformer layers (iter~1+ weighted), DiT ranges from 18\% to 86\% at iteration~0 before converging to low sparsity, and motion/dance transformer workloads split with EDGE low, MDM modest, and MLD high.  This per-layer variation motivates a layer-wise binary search for a threshold whose average hot fraction matches a target ratio $r$.  For layers with natural column sparsity (e.g., Stable Diffusion and Make-an-Audio transformer blocks), the calibrated thresholds settle within the physical activation range of 0.14--0.34, and the resulting cycle reduction reflects genuine column-skipping work that adapts to encoder-decoder asymmetry.  For layers without durable natural sparsity, calibration can degenerate.  DiT is the clearest example because late iterations naturally fall to low column sparsity, yet per-layer calibration pushes some thresholds as high as 1.64 and forces columns cold to meet the target hot ratio.  MDM and EDGE remain modest in the per-layer sweep and are therefore marked as calibration-limited cases when interpreting aggressive target ratios.

\subsection{Temporal Stability}
\label{sec:temporal}

\begin{figure}[t]
  \centering
  \includegraphics[width=\columnwidth]{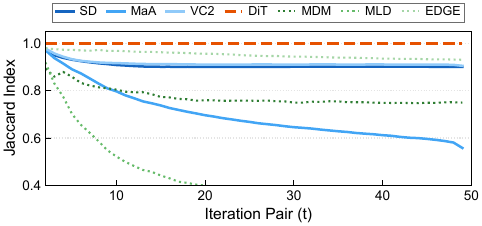}
  \caption{Jaccard stability across iterations.}
  \label{fig:jaccard}
\end{figure}

Figure~\ref{fig:jaccard} shows the Jaccard similarity index across iteration pairs for each model.  UNet+transformer workloads show high but imperfect stability, with mean Jaccard values of 0.906 for Stable Diffusion, 0.697 for Make-an-Audio, and 0.914 for VideoCrafter2.  DiT achieves Jaccard~$=$~1.0 despite its declining sparsity.  While the number of hot columns increases over time, no column that was hot ever becomes cold.

MDM (mean 0.774) and EDGE (mean 0.950) show moderate-to-high Jaccard, indicating that their small cold-column sets are relatively stable across iterations.  MLD instead shows the lowest stability of any model (mean 0.433, minimum 0.197), indicating that its substantial column sparsity churns substantively from one iteration to the next.  Two model factors explain this churn.  With only $M{=}6$ tokens per column, a single token crossing the threshold is enough to flip a column from cold to hot, and the $4\times$ expansion ratio leaves many borderline columns whose rank can swap between iterations.  The layout implication is that a static hot-cold partition computed once and reused across all iterations is a poor fit for MLD, even though it is more viable for the more stable models above.  Section~\ref{sec:perlayer_sweep} revisits this trade-off when comparing uniform and per-layer layouts on the cycle-level simulator.

\begin{figure}[h]
  \centering
  \includegraphics[width=\columnwidth]{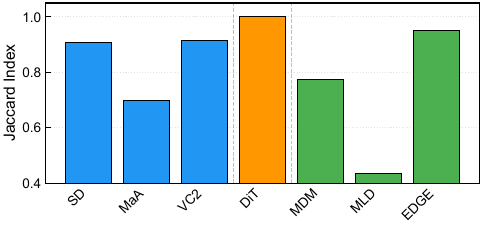}
  \caption{Mean Jaccard stability by group.}
  \label{fig:summary_arch}
\end{figure}

Figure~\ref{fig:summary_arch} summarizes the mean Jaccard stability index for all seven models, grouped by workload regime.  The three UNet+transformer workloads cluster at the high end (0.70--0.91), DiT reaches 1.0, and the motion/dance transformer workloads span the widest range, from EDGE at 0.950 down to MLD at 0.433.  The figure reinforces the three-way taxonomy.  Workload group and model dimensions determine not only how much column sparsity a model exhibits but also how stable that sparsity pattern is across denoising iterations.

\section{Memory System Evaluation}
\label{sec:evaluation}
\begin{table}[t]
\centering
\caption{Baseline simulation results.}
\label{tab:baseline}
\small
\begin{tabular}{lrccr}
\hline
Model & Ticks (B) & Compute & Stall & RBHR \\
\hline
DiT-XL/2      & 14.04 &  8.6\% & 88.7\% & 98.1\% \\
SD v1.4        &  9.99 & 11.6\% & 84.6\% & 98.6\% \\
VC2            & 53.04 & 11.9\% & 84.1\% & 98.6\% \\
MaA            &  1.45 & 11.6\% & 84.1\% & 99.2\% \\
MDM            &  0.40 & 11.2\% & 84.5\% & 99.7\% \\
MLD            &  0.02 &  8.7\% & 87.9\% & 99.7\% \\
EDGE           &  6.59 & 10.9\% & 84.7\% & 99.3\% \\
\hline
\end{tabular}
\end{table}
\subsection{Compute vs.\ Memory Stall Breakdown}

We validate the profiling-based taxonomy through cycle-level simulation under three layouts, including baseline row-major, uniform hot-cold at $\tau$, and per-layer hot-cold at target ratio $r$.  Table~\ref{tab:baseline} shows the execution time breakdown for all seven models under the baseline (all-dense, row-major) layout.  Memory stalls dominate across the board, comprising 84--89\% of total execution cycles.  Compute utilization ranges from 8.5\% (DiT) to 11.9\% (VideoCrafter2), indicating that the systolic array is idle for the vast majority of each iteration while waiting for data.  Row buffer hit rates are uniformly high (98.1--99.7\%), indicating that even the baseline layout achieves reasonable row locality.  The remaining row conflicts and misses nevertheless account for the dominant stall fraction because each conflict incurs a precharge-activate penalty of tens of nanoseconds, and these penalties accumulate over millions of DRAM accesses per iteration.  The absolute tick counts span three orders of magnitude, from 0.02 billion (MLD) to 53.04 billion (VideoCrafter2), reflecting differences in model size and token dimension.  Despite this scale variation, the compute-to-stall ratio is remarkably consistent across all seven models, which confirms that the memory bottleneck is structural to the FFN-Reuse dataflow rather than an artifact of any particular model's dimensions.  This consistency also means that even a small percentage-point improvement in row-buffer hit rate can yield substantial absolute cycle savings, particularly for the larger models.

\subsection{Uniform Threshold Sweep}
\label{sec:uniform_sweep}

\begin{figure}[t]
  \centering
  \includegraphics[width=\columnwidth]{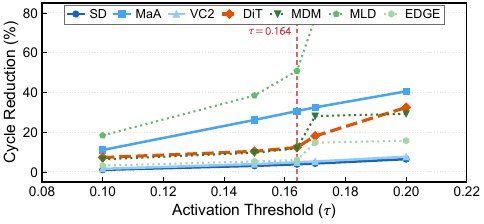}
  \caption{Cycle reduction vs.\ threshold $\tau$.}
  \label{fig:uniform_sweep}
\end{figure}

Figure~\ref{fig:uniform_sweep} plots cycle reduction versus activation threshold $\tau$ under the uniform hot-cold layout, with the $x$-axis sweeping $\tau$ from 0.10 to 0.20 and the $y$-axis reporting the percentage of baseline cycles eliminated.  UNet+transformer reductions differ substantially by workload at $\tau{=}0.164$.  Make-an-Audio reaches 30.6\%, while Stable Diffusion and VideoCrafter2 remain at 3.9\% and 4.9\%.  The disparity within this workload group traces back to the column-sparsity characterization in Section~\ref{sec:characterization}.  Make-an-Audio's smaller token dimension ($M{=}200$--$800$) and higher per-layer column sparsity produce more cold columns that the layout can group, whereas Stable Diffusion and VideoCrafter2 have lower column sparsity at this threshold despite their high element sparsity.  Activation concentration is therefore necessary for a static layout but does not by itself guarantee cycle benefit.  The cold-column count must also be large enough to improve row-buffer locality measurably.

Pure transformer DiT shows reductions of 7.3--32.5\% across the sweep, consistent with its dispersion profile in which column sparsity declines from ${\sim}$43\% to ${\sim}$10\% over 50 iterations.  At $\tau{=}0.164$, the reduction is 12.4\%.  The layout helps during early iterations when sparsity is higher, but returns diminish as dispersion progresses and more columns become active.  Motion/dance transformer workloads also split by workload.  MDM and EDGE achieve 11.9\% and 6.0\% at $\tau{=}0.164$, rising to 28.1--29.2\% and 14.8--15.7\% at $\tau{=}0.17$--$0.20$, while MLD is the strongest uniform-layout case, achieving 50.8\% at $\tau{=}0.164$ and up to 78.6\% at $\tau{=}0.20$.  MLD's extreme reduction follows directly from its model dimensions.  $M{=}6$ and $4\times$ expansion ratio produce the highest column sparsity of any model, and the layout groups a large fraction of columns into the cold partition.  These more aggressive thresholds must be interpreted together with the accuracy sweep, especially for motion workloads where quality degrades sharply above the reference point.

\begin{figure}[h]
  \centering
  \includegraphics[width=\columnwidth]{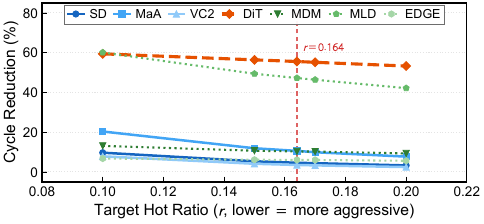}
  \caption{Per-layer reduction vs.\ ratio $r$.}
  \label{fig:perlayer_sweep}
\end{figure}

\subsection{Per-Layer Calibration Sweep}
\label{sec:perlayer_sweep}

Figure~\ref{fig:perlayer_sweep} shows cycle reduction versus target hot ratio $r$ under the per-layer hot-cold layout.  The $x$-axis represents the target hot ratio, with lower $r$ indicating a more aggressive setting, which is not directly comparable to the activation threshold $\tau$ in the uniform sweep.  UNet+transformer per-layer reductions remain workload-dependent.  Stable Diffusion ranges from 3.3--9.7\%, Make-an-Audio from 7.7--20.4\%, and VideoCrafter2 from 2.5--7.9\%.  At $r{=}0.164$, reductions are 4.6\%, 10.6\%, and 3.5\%, respectively.
Pure transformer DiT achieves 53.3--59.4\% under per-layer calibration, a substantial improvement over its 7.3--32.5\% uniform range.  However, 77.7\% of this per-layer reduction is an inflation artifact because DiT's natural column sparsity declines to ${\sim}$10\% by late iterations, but per-layer calibration forces most columns cold even in those iterations.  Motion/dance transformer per-layer results are similarly mixed.  MDM and EDGE remain modest at 9.3--13.1\% and 5.6--6.8\%, while MLD remains high at 42.2--60.1\%.  The MDM and EDGE values at low $r$ remain calibration-limited and should not be read as the same kind of genuine high-sparsity opportunity as MLD.  These cases illustrate why per-layer results must be interpreted alongside the calibrated thresholds and the underlying sparsity profile.

\subsection{Layout Sensitivity by Workload Group}

\begin{figure}[t]
  \centering
  \includegraphics[width=\columnwidth]{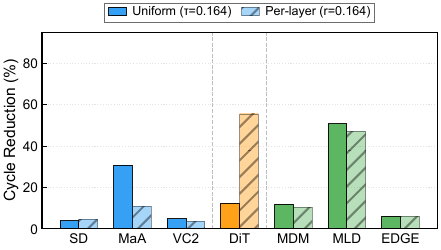}
  \caption{Layout sensitivity at $\tau{=}0.164$ and $r{=}0.164$.}
  \label{fig:layout_sensitivity}
\end{figure}

Figure~\ref{fig:layout_sensitivity} compares uniform and per-layer cycle reductions side by side at the primary operating point ($\tau{=}0.164$ / $r{=}0.164$) for all seven models, with paired bars for each model showing the two layout strategies.  The uniform layout tracks the profiling-based taxonomy.  Models with high column sparsity (MaA, MLD) show the strongest reductions, models with low column sparsity (SD, VC2, EDGE) show minimal benefit, and DiT and MDM fall in between.  Per-layer calibration substantially increases DiT's reported reduction (from 12.4\% to 53.3--59.4\%), but as noted in Section~\ref{sec:perlayer_sweep}, 77.7\% of that gain is an inflation artifact rather than genuine sparsity exploitation.  For UNet+transformer workloads, per-layer calibration provides only modest improvement over the uniform layout (e.g., 4.6\% vs.\ 3.9\% for SD), suggesting that the additional complexity of per-layer threshold tuning is not justified for this workload group at the primary operating point.  The layout sensitivity result therefore reinforces the need to combine cycle reduction with sparsity-profile and threshold-inflation analysis when evaluating memory layout strategies.

\subsection{Accuracy--Efficiency Tradeoff}

\begin{figure}[t]
  \centering
  \includegraphics[width=\columnwidth]{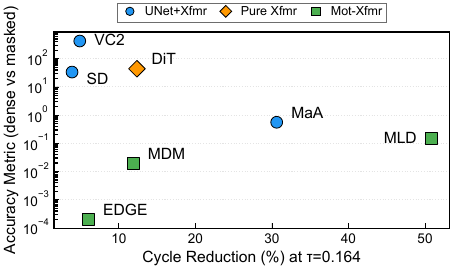}
  \caption{Accuracy vs.\ cycle reduction.}
  \label{fig:accuracy}
\end{figure}

Figure~\ref{fig:accuracy} and Table~\ref{tab:accuracy} present dense-versus-masked accuracy gaps and cycle reduction at $\tau{=}0.164$ for all seven models under the uniform layout ($N{=}100$ samples per model). Cycle reduction is computed as $(C_\mathrm{dense}-C_\mathrm{masked})/C_\mathrm{dense}$ relative to the dense baseline.
UNet+transformer workloads show workload-dependent tradeoffs.  MaA achieves the highest cycle reduction with moderate quality loss, while SD and VC2 show small reductions with correspondingly small metric deltas.  DiT yields a less favorable tradeoff per percentage point of reduction, consistent with its dispersion regime.  Among motion/dance transformer workloads, MLD stands out as the strongest layout case with 50.8\% cycle reduction and small motion FID degradation, while MDM and EDGE achieve modest reductions with minimal quality impact.

\begin{table}[h]
\centering
\caption{Accuracy gap and cycle reduction at $\tau{=}0.164$.}
\label{tab:accuracy}
\small
\setlength{\tabcolsep}{3pt}
\begin{tabular}{llrl}
\hline
Model & Quality & Dense vs. & Cycle  \\
Name& Metric& Masked Gap & Red.\\
\hline
DiT-XL/2 & FID / LPIPS & 44.0 / 0.104 & 12.4\% \\
SD v1.4 & FID / LPIPS & 33.6 / 0.069 & 3.9\% \\
VC2 & FVD / LPIPS$_\text{f}$ & 422.9 / 0.661 & 4.9\% \\
MaA & FAD / PESQ & 0.56 / 2.34 & 30.6\% \\
MDM & MPJPE / mFID & 0.019 / 0.027 & 11.9\% \\
MLD & MPJPE / mFID & 0.007 / 0.145 & 50.8\% \\
EDGE & $\Delta$BAS / mFID$_\text{k}$ & 0.0002 / 0.0001 & 6.0\% \\
\hline
\end{tabular}
\end{table}

\begin{figure}[t]
  \centering
  \includegraphics[width=\columnwidth]{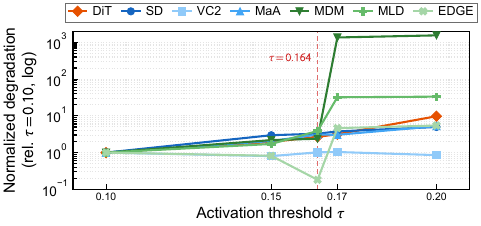}
  \caption{Accuracy degradation vs.\ threshold.}
  \label{fig:accuracy_sweep}
\end{figure}

Figure~\ref{fig:accuracy_sweep} shows accuracy degradation across all five thresholds, normalized to the $\tau{=}0.10$ value for cross-model comparison.  Together with Figure~\ref{fig:uniform_sweep}, the sweep motivates using $\tau{=}0.164$ as a primary reference point rather than as a globally optimal threshold.  It improves cycle reduction over lower thresholds while remaining immediately below the motion-model accuracy cliff between $\tau{=}0.164$ and $\tau{=}0.17$.  UNet+transformer workloads degrade smoothly across the sweep, whereas motion models enter a much more aggressive regime above the reference point.

Beyond the reference point, two regimes stand out.  DiT degrades steeply across the sweep, with FID rising $9.8\times$ from $\tau{=}0.10$ to $\tau{=}0.20$, consistent with dispersion causing aggressive thresholds to misclassify columns that remain active in later iterations.  Motion models exhibit an even sharper accuracy cliff between $\tau{=}0.164$ and $\tau{=}0.17$.  MDM's motion FID jumps $570\times$ and MLD's $8.3\times$, driven by column sparsity jumping from 12.1\% to 34.3\% (MDM) and 58.3\% to 88.0\% (MLD) at the higher threshold.  These observations support using $\tau{=}0.164$ as the primary reference point, where the threshold masks a limited set of genuinely cold columns rather than aggressively masking columns that affect output quality.

\section{Discussion}
\label{sec:discussion}
Our results show that workload group and column-level
sparsity jointly determine the effectiveness of memory-
layout optimization for diffusion accelerators. Element-level
sparsity, the metric most commonly reported in prior
diffusion-acceleration work, overestimates the exploitable
opportunity by up to 78 percentage points (Figure~\ref{fig:teaser}) and is therefore the wrong abstraction for memory-system design.  Column-level sparsity, evaluated at the tile granularity used by FFN-Reuse on systolic-array accelerators~\cite{heo2025exion}, predicts when a static hot-cold layout can recover row-buffer locality and when it cannot.  The three-way taxonomy from Section~\ref{sec:characterization} maps cleanly to three layout-design regimes discussed below, with the per-layer calibration caveat and the dominance of memory stalls (84--89\% of total cycles) framing the overall scale on which these layout choices matter. Also, the results indicate that sparsity magnitude alone is insufficient for evaluating diffusion acceleration opportunities. Two workloads with similar element-level sparsity can exhibit substantially different hardware
behavior once activations are viewed at column granularity.
Large token dimensions increase the probability that each
column contains at least one active token, collapsing the
amount of exploitable sparsity available to tiled hardware.
As a result, accelerator designs optimized primarily for
fine-grained activation sparsity may substantially overestimate the achievable reduction in memory traffic and execution time.

The taxonomy maps to a concrete decision procedure for accelerator designers, parameterized by workload group, token dimension $M$, and expansion ratio.  UNet+transformer workloads offer the clearest static-layout opportunity because the hot-cold column set stabilizes after the bootstrap iteration (Jaccard mean 0.70--0.91), so a one-time layout decision is structurally sound, but the designer should verify that enough cold columns exist to justify the layout complexity, since reductions range from 3.9\% (SD) to 30.6\% (MaA) at $\tau{=}0.164$.  Pure transformer workloads such as DiT exhibit perfect column-identity stability (Jaccard~$=$~1.0) despite declining sparsity, so an iteration-0 layout remains valid throughout generation, but the benefit is moderate (12.4\%) and diminishes as dispersion progresses.  Per-layer calibration inflates DiT's reduction to 53--59\%, though 77.7\% of that gain is an artifact of forcing columns cold in late low-sparsity iterations.  Motion/dance transformer workloads require checking $M$ and the expansion ratio.  Workloads with small $M$ and high expansion, such as MLD with $M{=}6$ and $4\times$ expansion, yield 50.8\% cycle reduction, while large-$M$ moderate-expansion workloads (MDM, EDGE) achieve only 6.0--11.9\% and are better served by element-level compute optimizations~\cite{heo2025exion}.  

The study also highlights an important architectural observation: under the FFN-Reuse execution model, diffusion inference remains fundamentally memory-bound. Across all seven workloads, memory stalls account for
84--89\% of total execution cycles despite already-high
baseline row-buffer hit rates. This implies that even modest
improvements in locality can translate into meaningful
end-to-end speedup, particularly for large diffusion models
with substantial off-chip activation traffic. Future diffusion
accelerators may therefore benefit from exposing sparsity-aware memory organization mechanisms directly within the
runtime system or memory controller rather than treating
sparsity solely as a compute-skipping opportunity.

Several caveats bound the scope of these findings.  The cycle-level simulator models heterogeneous UNet layers using a representative-block template at the $\text{ch}{=}320$ resolution level. Reduction ratios remain meaningful because the approximation applies equally to baseline and sparse modes, although absolute cycle counts carry a scaling factor. The study also evaluates only static layouts. Dynamic runtime repartitioning of hot and cold columns could improve efficiency for temporally unstable workloads such as MLD, although such schemes would introduce additional metadata overhead, control complexity, and data movement. Finally, all results are reported at a single DRAM operating point (GDDR6, 96\,GB/s). Different memory technologies or bandwidth points may shift the relative importance of layout optimization across workload groups.

\section{Related Work}
\label{sec:related}

\noindent\textbf{Diffusion models and modalities.}
Diffusion models began as iterative denoising generators~\cite{ho2020ddpm,song2021scorebased,nichol2021improved,dhariwal2021diffusion} and now span image~\cite{rombach2022latent,peebles2023dit,saharia2022imagen,ramesh2022hierarchical}, video~\cite{ho2022video,chen2024videocrafter2,blattmann2023svd}, audio~\cite{kong2020diffwave,chen2020wavegrad,huang2023makeanaudio}, and motion or dance synthesis~\cite{tevet2023mdm,chen2023mld,tseng2023edge}.  Although these works establish the diversity of diffusion workloads, none examine how the resulting differences in token dimension, hidden dimension, and expansion ratio affect memory-system behavior.  Our work fills this gap by characterizing column-level activation sparsity across three workload groups and four modalities.

\noindent\textbf{Diffusion acceleration (software).}
A large body of work reduces diffusion inference cost through fewer denoising steps or improved samplers~\cite{song2021ddim,lu2022dpmsolverpp,liu2022pndm,karras2022edm}, distillation~\cite{salimans2022progressive}, weight quantization~\cite{li2023qdiffusion,shang2023posttraining}, caching intermediate features~\cite{ma2024deepcache}, token merging~\cite{bolya2023tomesd}, or mobile-oriented model design~\cite{li2023snapfusion}.  These techniques are orthogonal to memory layout because they reduce the number or cost of iterations, while our characterization addresses how activation data is organized in memory within each iteration.  Importantly, software optimizations that reduce iteration count do not eliminate the memory-stall bottleneck within remaining iterations.  Our finding that 84--89\% of cycles are memory stalls suggests that layout optimization remains relevant even when fewer iterations are executed.

\noindent\textbf{Diffusion acceleration (hardware).}
EXION~\cite{heo2025exion} introduces FFN-Reuse on a systolic-array diffusion accelerator, reporting high FFN output sparsity and 52.47--85.41\% skipped FFN operations across its benchmarks.  These numbers are measured at element granularity and reported across only three workloads (DiT, Stable Diffusion, VideoCrafter2), which are all image or video models.  Our work extends the analysis in two ways.  We measure sparsity at the hardware-relevant column granularity, revealing that element-level numbers overstate exploitable sparsity by up to 78 percentage points, and we expand coverage to seven workloads across four modalities including audio and motion, where the sparsity profiles differ substantially.  Cambricon-D~\cite{kong2024cambricond} proposes diffusion-specialized hardware with a focus on temporal reuse across denoising steps.  Our characterization complements such designs by identifying which workload groups benefit from sparsity-aware data placement and which do not.

\noindent\textbf{DNN workload characterization.}
Workload characterization has long guided accelerator design, from benchmark suites such as MLPerf Inference~\cite{reddi2020mlperf} to recent IISWC studies of LLM inference workloads~\cite{cho2024llmservingsim} and transformer scaling behavior~\cite{pati2023twocs}.  These studies demonstrate the value of granularity-aware measurement.  LLM characterizations distinguish prefill from decode phases, and MLPerf separates offline from server scenarios.  Our work applies the same principle to diffusion models by distinguishing element-level from column-level sparsity and dense bootstrap from sparse denoising iterations, granularity distinctions that prior diffusion-acceleration work does not make.

\noindent\textbf{DNN, memory, and sparsity accelerators.}
General DNN accelerators~\cite{chen2014diannao,chen2014dadiannao,chen2016eyeriss,jouppi2017tpu,kwon2018maeri,parashar2019timeloop} and sparse-NN accelerators that exploit weight or input sparsity~\cite{han2016eie,zhang2016cambriconx,parashar2017scnn} motivate memory-aware layout design.  DRAM simulators~\cite{rosenfeld2011dramsim2,kim2016ramulator,luo2024ramulator2}, processing-in-memory~\cite{ham2016graphicionado,ahn2015tesseract}, sparse tensor/matrix~\cite{pal2018outerspace,hegde2019extensor,srivastava2020matraptor,zhang2021gamma}, and sparse GEMM accelerators~\cite{qin2020sigma,zhang2020sparch,yang2024trapezoid} explore locality-aware data organization.  These designs generally assume that the sparsity pattern is either static (e.g., pruned weights) or input-dependent but single-pass.  Diffusion activation sparsity differs in that the hot/cold partition is iteration-dependent, workload-specific, and may exhibit temporal drift (dispersion) or instability (churn) across the denoising loop, properties that our characterization quantifies for the first time.

\section{Conclusion}
\label{sec:conclusion}
This paper presented the first systematic characterization of column-level activation sparsity across seven diffusion workloads spanning three architectural regimes. Our results show that the element-level sparsity of 52--85\% reported in prior diffusion-acceleration work substantially overestimates
the sparsity that systolic-array accelerators can actually exploit, with gaps of up to 78 percentage points between
element-level and hardware-relevant column-level sparsity. The characterization reveals three distinct sparsity regimes. UNet+transformer workloads exhibit activation concentration that supports static layout optimization, while DiT exhibits dispersion as additional columns gradually become active over denoising iterations. Motion and dance transformer
workloads show mixed behavior governed primarily by token dimension $M$ and FFN expansion ratio. At $\tau{=}0.164$, column-level sparsity ranges from 3.9--30.6\% (UNet+transformer), 12.4\% (DiT), and 6.0--50.8\% (motion/dance transformer), with each group exhibiting a distinct temporal regime that determines static layout viability. Cycle-level simulation confirms that these distinctions translate to measurably different memory-system outcomes, with reductions of 3.9--50.8\% and memory stalls accounting for 84--89\% of total cycles. Column-level sparsity is therefore the actionable metric for diffusion accelerator memory-system design. 

Several opportunities remain for extending this work. Dynamic runtime remapping may further improve efficiency, while 
emerging architectures including flow-matching and consistency models may exhibit different sparsity dynamics that require revisiting the taxonomy introduced in this work. Future research could also explore co-design between column-level layout optimization and complementary techniques such as quantization, token merging, and 
near-memory acceleration.

\bibliographystyle{IEEEtranS}
\bibliography{reference}

%
%
%

\end{document}